\documentclass[twocolumn,aps,prl,epsfig,showpacs,superscriptaddress]{revtex4}

\usepackage{graphicx}

\begin{document}

\title{Spin filtering through ferromagnetic BiMnO$_3$ tunnel barriers}

\author{M. Gajek}
\affiliation{Unit\'e Mixte de Physique CNRS / Thales, Domaine de Corbeville, 91404 Orsay, France}
\affiliation{Institut de Ci\`encia de Materials de Barcelona, CSIC, Campus de la UAB, 08193 Bellaterra, Spain}
\author{M. Bibes}
\affiliation{Unit\'e Mixte de Physique CNRS / Thales, Domaine de Corbeville, 91404 Orsay, France}
\affiliation{Institut d'Electronique Fondamentale, Universit\' e Paris-Sud, 91405 Orsay, France}
\author{A. Barth\'el\'emy}
\email{agnes.barthelemy@thalesgroup.com}
\author{K. Bouzehouane}
\affiliation{Unit\'e Mixte de Physique CNRS / Thales, Domaine de Corbeville, 91404 Orsay, France}
\author{S. Fusil}
\affiliation{Unit\'e Mixte de Physique CNRS / Thales, Domaine de Corbeville, 91404 Orsay, France}
\affiliation{Universit\'e d'Evry, rue du P\`ere Jarlan, 91025 Evry, France}
\author{M. Varela}
\affiliation{Dept. de F\'{i}sica Aplicada i \`{O}ptica, Universitat de Barcelona, Diagonal 647, 08028 Barcelona,
Spain}
\author{J. Fontcuberta}
\affiliation{Institut de Ci\`encia de Materials de Barcelona, CSIC, Campus de la UAB, 08193 Bellaterra, Spain}
\author{A. Fert}
\affiliation{Unit\'e Mixte de Physique CNRS / Thales, Domaine de Corbeville, 91404 Orsay, France}

\date{\today}

\begin{abstract}

\vspace{0.5cm}

We report on experiments of spin filtering through ultra-thin single-crystal layers of the insulating and
ferromagnetic oxide BiMnO$_3$ (BMO). The spin polarization of the electrons tunneling from a gold electrode
through BMO is analyzed with a counter-electrode of the half-metallic oxide La$_{2/3}$Sr$_{1/3}$MnO$_3$ (LSMO).
At 3 K we find a 50\% change of the tunnel resistances according to whether the magnetizations of BMO and LSMO
are parallel or opposite. This effect corresponds to a spin filtering efficiency of up to 22\%. Our results thus
show the potential of complex ferromagnetic insulating oxides for spin filtering and injection.

\vspace{0.5cm}
\end{abstract}


\maketitle

Obtaining highly spin-polarized electron tunnelling is an important challenge in nowadays spintronics, either
for spin injection into semiconductors \cite{rashba2000,fert2002} or magnetoresistive effects \cite{julliere75}.
The classical way is by tunnelling from a ferromagnetic conductor through a non-magnetic barrier. This is the
basic mechanism of the tunnelling magnetoresistance (TMR) of tunnel junctions composed of two ferromagnetic
electrodes (spin emitter and spin analyzer) separated by a nonmagnetic insulator \cite{moodera95}. Such tunnel
junctions are currently applied to the development of sensors and memories (MRAM). Spin polarized tunnelling
from a ferromagnetic metal through a non-magnetic layer is also what can be used for spin injection into a
semiconductor \cite{motsnyi2003}. Another way for spin polarized tunnelling has been little explored: this is
tunnelling from a non-magnetic electrode through a ferromagnetic insulator. The concept was introduced by
Moodera et al \cite{moodera88} with EuS tunnel barriers. The effective barrier height of an insulating layer
corresponds to the energy difference between the Fermi level and the bottom of the conduction band (or the top
of the valence band). A spin dependent barrier height is therefore expected from the spin splitting of the
energy bands in a ferromagnetic insulator. The exponential dependence of the tunnelling on the barrier height
can lead to a very efficient spin filtering. This has been confirmed, at least at low temperature, by the very
high spin polarizations obtained by tunnelling through barriers of EuS and EuSe \cite{moodera88,moodera93} and
more recently with EuO \cite{santos2004}. Spin filtering tunnel barriers can be of high interest for spin
injection into semiconductors without using ferromagnetic metals as spin polarized injectors. Very large
magnetoresistance effects can also be expected by switching from parallel to antiparallel the magnetic
configuration of two spin filter barriers in a double junction \cite{worledge2000}.

To demonstrate spin filtering by a ferromagnetic barrier, the spin polarization of the current tunnelling from a
nonmagnetic electrode can be analyzed either with a superconductor \cite{moodera88,moodera93}, or with a
ferromagnetic counter-electrode \cite{leclair2002}. In the latter case, the ferromagnetic counter-electrode
collects differently the spins parallel and antiparallel to its magnetization, so that the current depends on
the relative orientations of the magnetic moments of the ferromagnetic barrier and counter-electrode. This is
illustrated by the experiments of LeClair et al \cite{leclair2002} with an Al electrode, an EuS barrier and a
counter-electrode of ferromagnetic Gd. A TMR of up to 130 \% at 2K has been obtained with this type of tunnel
junction \cite{leclair2002}.

Up to now, the only experiments of spin filtering by ferromagnetic barriers have been performed with insulating
layers of Eu chalcogenides. However, the very low Curie temperature of EuS (16 K) or EuSe (4.6 K), and the poor
chemical compatibility of the Eu chalcogenides with many possible electrode materials limit their practical
potential for spin filtering. The list of other possible candidates includes a few ferromagnetic perovskite
oxides and a large family of ferrites (spinels and garnets). Compared to the complex crystal structure of the
ferrites, perovskites are relatively simple and more convenient for integration into tunnel heterostructures,
particularly if an isostructural fully polarized half-metallic ferromagnetic metal, such as
La$_{2/3}$Sr$_{1/3}$MnO$_3$ (LSMO) \cite{bowen2003} is used as a spin analyzer to probe the filter efficiency.

BiMnO$_3$ (BMO) is an insulating and ferromagnetic perovskite oxide, having a Curie temperature (T$_C$) of 105 K
and a magnetic moment of 3.6 $\mu_B$/formula unit (in bulk) \cite{chiba97}. It is a highly insulating compound
and, remarkably, the insulating state is very robust \cite{chiba97}. Experimental determinations of the exchange
splitting of the (empty) conduction band of BMO have not been reported; however it can be estimated to about
~0.5 eV from linear spin-density approximation (LSDA) calculations \cite{hill99} and to ~1.6 eV from LSDA+U
\cite{shishidou2004}. In both cases, the gap is smaller for spin-up electrons, so that when used as a spin
filter barrier, a BMO layer should filter out spin-down electrons and produce a positively spin-polarized
current. From the gap found by LSDA+U, a computation technique which is commonly accepted to be more reliable to
calculate band gaps, it follows that the exchange splitting in BMO is larger than that predicted for EuS (0.36
eV \cite{wachter79}) and EuO (0.6 eV \cite{steeneken2002}), which should result in an increased spin-filtering
efficiency. Therefore, both from the electronic point of view and from materials perspective, BMO appears as an
ideal perovskite to be implemented as a spin-filter barrier.

In this Letter we report on the growth of thin epitaxial layers of the BiMnO$_3$ perovskite and their
integration in spin-filter structures. We demonstrate the spin-filtering properties of tunnel barriers of BMO in
Au-BMO-LSMO junctions. The device can be operated up to about 40K. Our results demonstrate the potential of
complex ferromagnetic oxides for high temperature spin filtering and spin injection.

\begin{figure}[!h]
 \includegraphics[keepaspectratio=true,width=\columnwidth]{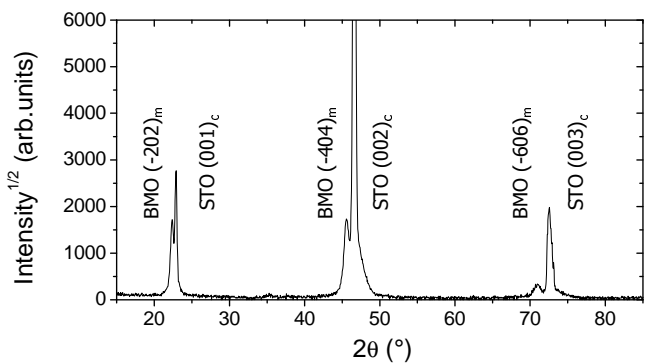}
 \caption{$\theta$-2$\theta$ scan of a 30 nm film grown at 625$^\circ$C.}
 \label{xrd}
\end{figure}

BMO thin films were prepared on (001) SrTiO$_3$ substrates by pulsed laser deposition using a KrF excimer laser
($\lambda$=248nm). The growth of BMO was carried out from a non-stoichiometric multiphase target with a Bi:Mn
ratio of 1.15, in an oxygen pressure of 0.1 mbar. Bulk BMO has a heavily distorted perovskite structure that can
be represented in the monoclinic C2 space group \cite{atou99}. In the triclinic pseudo-cubic unit cell the
lattice parameters are a=c=3.985 $\rm{\AA}$, b=3.989 $\rm{\AA}$ with $\alpha$=$\gamma$=91.4$^\circ$,
$\beta$=91$^\circ$ \cite{atou99}. Extensive details on film growth and structural characterization will be
reported elsewhere \cite{gajek2004}. Here we just mention that single-phase BMO films have only been obtained in
a narrow temperature window around 625$^\circ$C.

In figure \ref{xrd} we show a $\theta$-2$\theta$ scan of a BMO film of nominal thickness 30 nm. Diffraction
peaks occurring at slightly lower angles than the (00$l$)$_c$ reflections ($c$ : pseudocubic representation) of
the STO substrate are clearly visible and could be indexed as (0$l$0)$_c$ reflections of the BMO film.  They
correspond to ($l0l$)$_m$ in the monoclinic ($m$) system. We do not detect ($lll$)$_m$ and ($3lll$)$_m$
reflections, as found by Moreira dos Santos et al \cite{moreira2004}. $\phi$-scans of the (111)$_c$ reflections
of the BMO layer and STO substrate (not shown) indicate a cube-on-cube growth. The out-of-plane parameter (c)
deduced from the angular position of the (040)$_c$ reflection is 3.96 $\rm{\AA}$, close to the b parameter in
bulk (3.989 $\rm{\AA}$). As c is inferior to the bulk parameter in spite of the compressive strain induced by
mismatch of -0.7\% with the substrate, the reduction of the cell volume with respect to bulk is likely to be due
to some Bi deficiency.

\begin{figure}[!h]
 \includegraphics[keepaspectratio=true,width=0.9\columnwidth]{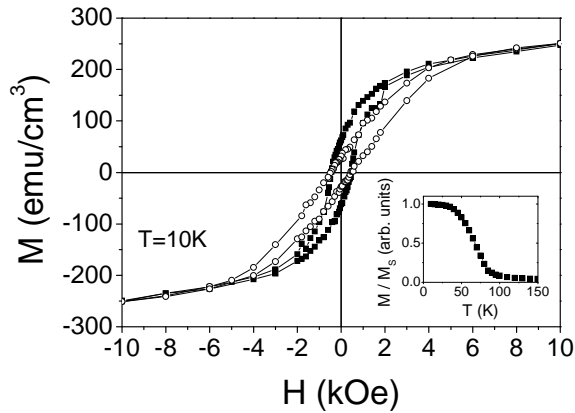}
 \caption{Magnetization hysteresis cycles measured at 10K with the field applied in-plane (solid symbols) and
 out-of-plane (open symbols). Inset : temperature dependence of the magnetization measured in a field of 1 kOe.}
 \label{magn}
\end{figure}

On figure \ref{magn}, we plot the magnetization (M) vs applied magnetic field (H) for a 30 nm thick BMO film
after subtracting the diamagnetic contribution of the STO substrate. We observe a clear ferromagnetic behavior
with a coercive field of ~470 Oe measured in-plane and out-of-plane, and a remanence of 62 emu/cm$^3$ with the
field in plane and 29 emu/cm$^3$ out-of-plane. The shape of the magnetization loops indicates that the easy axis
clearly lies in the film plane while the out-of-plane direction is a hard axis. The magnetization is not
saturated even in a field of several teslas. It reached only reaching 280 emu.cm$^{-3}$ at 5T, and is thus
fairly reduced with respect to the bulk \cite{chiba97} (M(5T) $\simeq$ 0.52 M$_{S \hspace{0.2em} bulk}$), which
is consistent with the results of Ohshima et al \cite{ohshima2000}. The slow increase of the magnetization at
high field is likely to result from the progressive realignment of canted spins. Both the low magnetization and
this canted behavior could be explained by the presence of Bi vacancies which locally disturb the complex
orbital ordering essential for the long-range ferromagnetic order in BMO \cite{moreira2002}. The temperature
dependence of the magnetization of this 30 nm film (see inset of Figure \ref{magn}) indicates that the
ferromagnetic transition occurs in the vicinity of 97K, which is close to the bulk value ($\sim$105K).

We have measured the temperature dependence of the resistivity of a 30 nm BMO film in the 150-300K range and
found a thermally activated behavior with a room-temperature resistivity of $\rho_{300K}$=175$\Omega$cm
($\rho_{300K}$ =20 k$\Omega$cm for bulk \cite{chiba97}) and an activation energy of E$_a$=239 meV (E$_a$=262 meV
for bulk \cite{chiba97}). Below 150K, the film resistance was exceedingly large to be measured with the
available experimental set up. Using the room-temperature resistivity value and the activation energy we
estimate the resistivity around T$_C$ to about 5 G$\Omega$cm. This value, somewhat smaller than that of bulk BMO
ceramics but similar to what is reported for Bi$_{0.9}$Sr$_{0.1}$MnO$_3$ \cite{chiba97}, is large enough for the
BMO film to be taken as a good insulator.

\begin{figure}[!h]
 \includegraphics[keepaspectratio=true,width=\columnwidth]{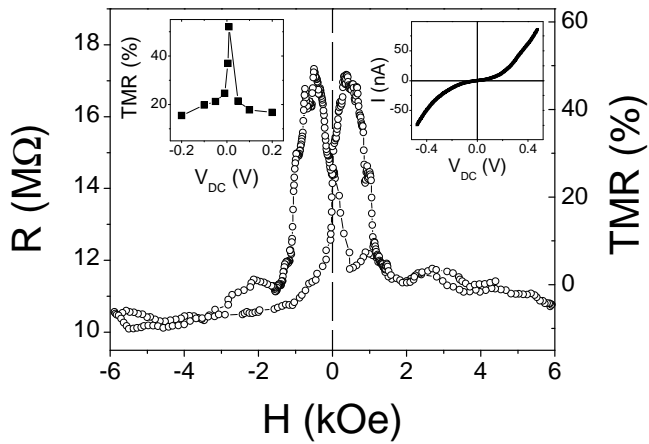}
 \caption{Field dependence of the resistance of a junction at 3K (V$_{DC}$=10 mV). Insets : bias dependence of the TMR (left); I(V) curve of the
 junction (right).}
 \label{tmr}
\end{figure}

In order to probe the potential of BMO as a ferromagnetic barrier for spin filtering, ultrathin BMO films (3.5
nm) were grown onto a STO (1nm)/LSMO(25nm)//STO template. The intercalated 1 nm of STO layer is to magnetically
decouple the BMO barrier from the LSMO electrode. One also knows that the half-metallic character of LSMO is
conserved at the interface with STO \cite{bowen2003,garcia2004}. Atomic force microscopy (AFM) images of this
structure show a very smooth surface suitable for patterning the sample into tunnel junctions with the following
structure: Au/BMO/STO/LSMO.

Small junctions (50nm x 50nm) were patterned by a nanolithography process based on the indentation of thin
resist by conductive-tip AFM followed by the filling of the resulting hole with a sputtered Au layer
\cite{bouzehouane2003}. In these experiments, the resistance of the LSMO bottom electrode was always small
enough to ensure homogeneous current flow through the junction.  The I(V) curve of the right inset in figure
\ref{tmr} exhibits clearly the non-linear and asymmetric behavior expected for tunnel junctions with different
electrodes.

The R(H) plot of a Au/BMO/STO/LSMO junction in Fig. \ref{tmr} is typical of TMR curves with a TMR of about 50\%.
The sharp increase of resistance at small field corresponds to the magnetic reversal of LSMO at its coercive
field of about 100 Oe. The resistance drops back to its low-level value above 1.5 kOe, which is close to the
value at which the magnetization cycle of the 30 nm BMO film closes (see figure \ref{magn}). The resistance
maximum corresponds to the antiparallel configuration of the magnetization of LSMO with the remanence of BMO
($\sim$25\% of saturation). The slow and almost linear resistance variation at fields above 2kOe is expected
from the high-field susceptibility observed in the M(H) cycles (see figure \ref{magn}). A part of this variation
might also be due to reorientation of canted spins at the LSMO/STO interface \cite{garcia2004}.

The positive value of the TMR is in agreement with the calculated band structure of BMO
\cite{hill99,shishidou2004}. Using an extension of the Julli\`ere model \cite{julliere75}
(TMR=2P$_1$P$_2$/(1-P$_1$P$_2$), where P$_1$=90\% is the typical spin-polarization of LSMO at the interface with
STO \cite{garcia2004} and P$_2$ the spin-polarization due to the BMO spin filter effect), the measured TMR=50\%
corresponds to a spin-filter polarization of 22\%. However, as the magnetization of the BMO film at the reversal
field of LSMO is only 25 \% of its saturation value, we can renormalize the spin-filter polarization to 88 \%.
This value is close to the maximum spin-filter polarization found for EuS ($\sim$85\%) \cite{moodera88}, but
still lower than expected from the calculated value of the exchange splitting.

As shown in the inset (left) of figure \ref{tmr}, the TMR decreases at increasing bias. This feature is common
in MTJs \cite{han2001} and ascribed in large part to magnon excitations at the electrode-barrier interfaces
\cite{zhang97}. This mechanism is certainly also active here on the LSMO interface, but, since only one of the
electrodes is magnetic, it cannot account for an approximately equal drop in positive and negative bias. A
symmetric drop can only be due to magnon excitations \emph{inside} the BMO barrier. With a tunnelling current
predominantly carried by electrons having a complex momentum component perpendicular to the layers and zero
parallel component, excitations of magnons of parallel momentum can flip the spin of these electrons and scatter
them into evanescent waves of different decay length. This can affect strongly the conductance and the TMR.
Although this magnon contribution to the bias dependence of the TMR should be more important in spin filters
than in conventional MTJs, they have not been incorporated in the existing spin filter models
\cite{saffarzadeh2004}, and certainly deserves the attention of theorists.

\begin{figure}[!h]
 \includegraphics[keepaspectratio=true,width=\columnwidth]{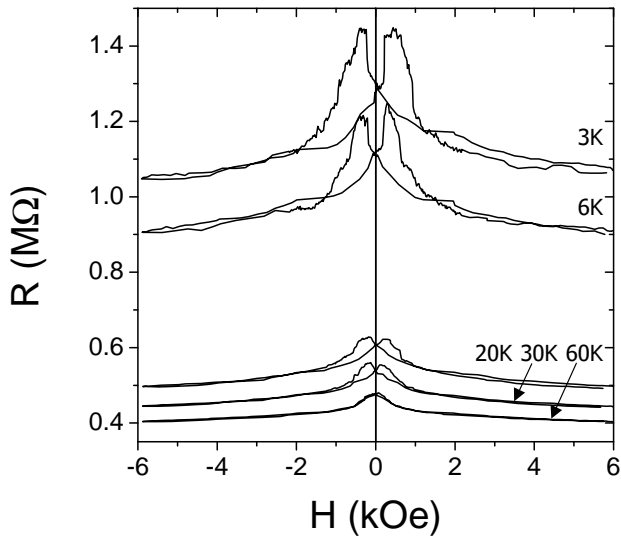}
 \caption{Field dependence of the resistance at different temperatures for a second junction (V$_{DC}$=10 mV).}
 \label{tmr(t)}
\end{figure}

In figure \ref{tmr(t)} we plot R(H) curves obtained for another Au/BMO/STO/LSMO junction at different
temperatures. At 3K, its resistance is somewhat lower than that of the junction of Fig. \ref{tmr}, which might
be due to a slightly lower barrier thickness. The TMR of this junction is 29\% at 3K and then gradually
decreases at increasing temperature. Beyond 40K, it remains only a small and reversible variation that should be
predominantly due to spin canting reorientation. The temperature at which the spin-filter effect vanishes is
thus lower than the Curie temperature of our 30 nm BMO films (see inset of figure \ref{magn}). This may indicate
that the T$_C$ of BMO ultrathin layers is depressed compared to bulk value. It is also possible that when
temperature increases, the magnetization of the BMO barrier becomes increasingly coupled to that of the LSMO
electrode, so that an antiparallel configuration can no longer be obtained. Further work is required to clarify
this point.

In summary, we have grown single-phased thin films of the ferromagnetic insulator BiMnO$_3$ on (001)-oriented
SrTiO$_3$ substrates. Spin filtering by a BMO tunnel barrier has been demonstrated by magneto-transport
measurements on Au-BMO-LSMO junctions which have shown up to 50\% of TMR. The TMR decreases rapidly and
symmetrically as a function of the bias voltage, which can be the signature of magnon excitations inside the
magnetic barrier. This new inelastic scattering mechanism was not included in the theory of spin filter
junctions \cite{saffarzadeh2004} and has to be studied in more detail. Our results suggest that BMO could be
used for spin-injection into semiconductors as high-quality perovskite/Si \cite{mckee98} and perovskite/GaAs
\cite{liang2004} structures have already been fabricated. Further work is needed to fully understand and improve
the magnetic properties of BMO ultrathin film but this is the first experimental evidence of spin filtering with
a complex oxide and thus constitutes a hallmark towards spin-filters operating at room temperature, using spinel
ferrites for instance. In addition, since BMO is also ferroelectric \cite{son2004} and as a coupling between the
magnetic and dielectric properties in this material has been recently reported \cite{kimura2003}, our experiment
can be thought as a preliminary stage in the exploitation of multiferroic materials in spintronics devices.

\acknowledgements{This work has been supported in part by the MCyT (Spain) projects MAT2002-03431, FEDER, the
Franco-Spanish project HF-20020090 and the E.U. STREP "Nanotemplates" (Contract number: NMPA4-2004-505955). M.G.
acknowledges financial support from ICMAB, through the Marie Curie Training Site program.}

\vspace{0.5em}

\bibliographystyle{prsty}

\end{document}